\documentclass[longauth]{aa}  

\usepackage{graphicx}
\usepackage{txfonts}
\usepackage[colorlinks = true,
            linkcolor = blue,
            urlcolor  = blue,
            citecolor = blue,
            anchorcolor = blue,
            pdfencoding = auto]{hyperref}

\usepackage{booktabs}
\usepackage{caption}
\usepackage{subcaption}
\usepackage[flushleft]{threeparttable}
\usepackage{lipsum}
\usepackage{balance}

\makeatletter
\renewcommand*\aa@pageof{
, page \thepage{} of \pageref*{LastPage}}
\makeatother

\begin{document}

\titlerunning{Molecular gas clumpiness under the influence of AGN}
\authorrunning{Esposito et al.}

\title{
Galaxy Activity, Torus and Outflow Survey (GATOS) X: \\
Molecular gas clumpiness under the influence of AGN
}

\author{
Federico Esposito \inst{1} \and
Almudena Alonso-Herrero \inst{2} \and
Santiago García-Burillo \inst{1} \and
Ismael García-Bernete \inst{2} \and
Françoise Combes \inst{3, 4} \and
Richard Davies \inst{5} \and
Enrique Lopez-Rodriguez \inst{6, 7} \and
Omaira González-Martín \inst{8} \and
Cristina Ramos Almeida \inst{9, 10} \and
Anelise Audibert \inst{9, 10} \and
Erin K. S. Hicks \inst{11} \and
Miguel Querejeta \inst{1} \and
Claudio Ricci \inst{12, 13} \and
Enrica Bellocchi \inst{14, 15} \and
Peter Boorman \inst{16} \and
Andrew J. Bunker \inst{17} \and
Steph Campbell \inst{18} \and
Daniel E. Delaney \inst{11} \and
Tanio Díaz-Santos \inst{19, 20} \and
Donaji Esparza-Arredondo \inst{8} \and
Sebastian Hönig \inst{21} \and
Álvaro Labiano Ortega \inst{22} \and
Nancy A. Levenson \inst{23} \and
Chris Packham \inst{24} \and
Miguel Pereira-Santaella \inst{25} \and
Rogemar A. Riffel \inst{26, 27} \and
Dimitra Rigopoulou \inst{17, 19} \and
David J. Rosario \inst{18} \and
Antonio Usero \inst{1} \and
Lulu Zhang \inst{21}
}

\institute{
Observatorio de Madrid, OAN-IGN, 
Alfonso XII, 3, E-28014 Madrid, Spain\\
\email{\href{mailto:f.esposito@oan.es}{f.esposito@oan.es}}
\and
Centro de Astrobiología (CAB, CSIC-INTA), 
Camino Bajo del Castillo s/n, 
E-28692 Villanueva de la Cañada, Madrid, Spain
\and
Observatoire de Paris, LUX, PSL University, 
Sorbonne Université, CNRS, F-75014 Paris, France
\and 
Collège de France, 11 Place Marcelin Berthelot, 75231 Paris, France
\and 
Max-Planck-Institut für Extraterrestrische Physik, 
Postfach 1312, 85741, Garching, Germany
\and 
Department of Physics \& Astronomy, University of South Carolina, 
Columbia, SC 29208, USA
\and 
Kavli Institute for Particle Astrophysics \& Cosmology (KIPAC), 
Stanford University, Stanford, CA 94305, USA
\and 
Instituto de Radioastronomía y Astrofísica (IRyA-UNAM),
Antigua Carretera a Pátzcuaro \#8701, 
Ex-Hda. San José de la Huerta, C.P. 58089, Morelia, Michoacán, Mexico
\and 
Instituto de Astrofísica de Canarias (IAC), Calle Vía Láctea, s/n,
38205 La Laguna, Tenerife, Spain
\and 
Departamento de Astrofísica, Universidad de La Laguna, 38206 La
Laguna, Tenerife, Spain
\and 
Department of Physics \& Astronomy, University of Alaska Anchorage, 
Anchorage, AK, 99508-4664, USA
\and 
Department of Astronomy, University of Geneva, 
1290 Versoix, Switzerland
\and 
Instituto de Estudios Astrofísicos, Facultad de 
Ingeniería y Ciencias, Universidad Diego Portales, 
Av. Ejército Libertador 441, Santiago, Chile
\and 
Departamento de Física de la Tierra y Astrofísica, 
Fac. de CC. Físicas, Universidad Complutense de Madrid, 
28040 Madrid, Spain
\and 
Instituto de Física de Partículas y del Cosmos IPARCOS, 
Universidad Complutense de Madrid, 28040 Madrid, Spain
\and 
Cahill Center for Astrophysics, California 
Institute of Technology, 1216 East California 
Boulevard, Pasadena, CA 91125, USA
\and 
Department of Physics, University of Oxford, Denys Wilkinson Building, 
Keble Road, Oxford, OX13RH, UK
\and 
School of Mathematics, Statistics and Physics, 
Newcastle University, Newcastle upon Tyne, 
NE1 7RU, UK
\and 
Institute of Astrophysics, Foundation for Research and 
Technology -- Hellas (FORTH), Voutes 70013, Heraklion, Greece
\and 
School of Sciences, European University Cyprus, Diogenes street,
Engomi 1516 Nicosia, Cyprus
\and 
School of Physics \& Astronomy, University of Southampton, 
Southampton, SO17 1BJ, Hampshire, UK
\and 
Telespazio UK for the European Space Agency (ESA), ESAC,
Camino Bajo del Castillo s/n, 28692 Villanueva de la Cañada,
Madrid, Spain
\and 
Space Telescope Science Institute, 
3700 San Martin Drive Baltimore, MD 21218, USA
\and 
Department of Physics and Astronomy, The University of Texas at San Antonio, 
1 UTSA Circle, San Antonio, TX 78249, USA
\and 
Instituto de Física Fundamental, CSIC, 
Calle Serrano 123, 28006, Madrid, Spain
\and 
Centro de Astrobiología (CAB, CSIC/INTA),
Ctra de Torrejón a Ajalvir, km 4,
28850 Torrejón de Ardoz, Madrid, Spain
\and 
Departamento de Física, CCNE, Universidade Federal de Santa Maria, 
Av. Roraima 1000, 97105-900, Santa Maria, RS, Brazil
}

\date{Received XXX; accepted YYY}

  \abstract{
  The distribution of molecular gas on small scales 
  regulates star formation and the growth of supermassive 
  black holes in galaxy centers, yet the role of 
  active galactic nuclei (AGN) feedback in shaping 
  this distribution remains poorly constrained.   
  We investigate how AGN
  influence the small-scale structure of molecular gas 
  in galaxy centers, by measuring the clumpiness of 
  CO($3-2$) emission observed with the Atacama 
  Large Millimeter/submillimeter Array (ALMA)
  in the nuclear regions ($50-200$ pc from the AGN) of 
  16 nearby Seyfert galaxies from the
  Galaxy Activity, Torus, and Outflow Survey (GATOS).
  To quantify clumpiness, we apply three different methods: 
  (1) the median of the pixel-by-pixel contrast between 
  the original and smoothed maps; (2) the ratio of the total
  excess flux to the total flux, after substracting the
  background smoothed emission; and (3) the fraction 
  of total flux coming from clumpy regions,
  interpreted as the mass fraction in clumps.
  We find a negative correlation between molecular gas 
  clumpiness and AGN X-ray luminosity ($L_{\text{X}}$), 
  suggesting that higher AGN activity is associated 
  with smoother gas distributions.
  All methods reveal a turnover in this relation
  around $L_{\text{X}} = 10^{42}$ erg s$^{-1}$,
  possibly indicating a threshold above which AGN feedback 
  becomes efficient at dispersing dense molecular structures
  and suppressing future star formation.
  Our findings provide new observational evidence that 
  AGN feedback can smooth out dense gas structures 
  in galaxy centers.
  }

   
   \maketitle

\section{Introduction}

Active galactic nuclei (AGN) are powerful sources 
of energy and momentum that can significantly 
impact the interstellar medium (ISM) 
of their host galaxies.
The feedback processes driven by AGN, 
including both radiative and mechanical effects,
play a crucial role in shaping the gas distribution, 
star formation, and overall evolution of galaxies
\citep{fabian12, kormendy13, heckman14, morganti17, harrison18}.
Deciphering the mechanisms by which AGN feedback 
affects the ISM and drives galaxy evolution
is one of the major challenges
in modern astrophysics
\citep[see][for a recent review]{harrison24}.

Within a galaxy,
molecular gas can organize into
Giant Molecular Clouds (GMCs), 
with typical masses
$M_{\text{GMC}} \sim 10^4 - 10^6$ M$_{\odot}$
and sizes of $\sim 10 - 50$ pc 
\citep{solomon87, omont07, fukui10, heyer15,
chevance20, chevance23}.
Such clouds are sufficiently massive
and dense that their self-gravity
tends to drive gravitational collapse.
However, GMCs can achieve a quasi-stable
state because other internal forces
-- mainly magnetic pressure and turbulent motions --
provide support that counteracts collapse
\citep{mckee07, crutcher12}.
Within them there exists a complex hierarchical 
structure of smaller and smaller
high density structures
\citep{elmegreen96, federrath12, elia18, buck22},
which are thought to be the sites of star formation
\citep{bergin07, heyer15, massi19}.

AGN-driven radiation, winds, jets, and cosmic rays
can further compress, evaporate, or disperse the gas
\citep{pier95, schartmann11, namekata14, pozzi17, 
vallini17, mingozzi18, circosta21, 
gabici22, bertola24, koutsoumpou25},
and although there are hints of positive feedback
\citep[e.g.][]{maiolino17, bessiere22, mercedesfeliz23},
we mostly observe molecular gas
depletion in the central regions of active galaxies
\citep{sabatini18, rosario19, ellison21b, 
garciaburillo21, garciaburillo24}.
Among the clearest signatures of AGN impact on the host ISM,
\citet{garciaburillo21, garciaburillo24} reported a systematic
deficit in the central concentration of cold molecular gas 
in AGN-host galaxies, quantified through the concentration 
index $\log (\Sigma_{50} / \Sigma_{200})$, 
where $\Sigma_{50}$ and $\Sigma_{200}$ are the 
average CO surface densities within circular apertures 
of 50 pc and 200 pc radii, respectively.
They found that this concentration index shows a turnover 
as a function of AGN X-ray luminosity, with a change
around $L_{\text{X}} (2-10 \text{ keV}) \sim 10^{41.5}$
erg s$^{-1}$.
While the concentration index quantifies how centrally 
peaked the gas distribution is, it does not capture 
the small-scale structure within the nuclear region.
In this work, we focus on the clumpiness 
of the molecular gas, a property that reflects 
how fragmented or structured the emission is
within the nuclear region.

Molecular gas clumpiness, characterised by an 
irregular and dense distribution, 
can serve as an indirect measure 
of star formation activity, 
as the most clumpy regions are where the
gravitational collapse leads to the formation of new stars
\citep{krumholz14, krumholz19}.
Understanding the potential role that AGN feedback 
plays in shaping the molecular gas clumpiness 
of their host galaxies is crucial
for assessing how AGN can regulate star 
formation and the evolution of galaxy centers.
To this end,
this work is part of the 
Galaxy Activity, Torus, and Outflow Survey 
\citep[GATOS\footnote{
\url{https://gatos.myportfolio.com}},][]{
garciaburillo21, alonsoherrero21,
garciabernete24, garciabernete24b, zhang24, 
poitevineau25, fuller25},
an ongoing effort to study the properties of AGN,
of the dusty molecular tori, and the AGN interaction 
with the host galaxy in local Seyfert galaxies 
using high resolution observations.

The paper is structured as follows:
in Section~\ref{sec:sample} we introduce the sample
of galaxies, in Section~\ref{sec:analysis} we describe
the methods for measuring clumpiness, 
and the statistical tools employed in our analysis.
Section~\ref{sec:results} presents the results
of our analysis, and in Section~\ref{sec:discussion}
we discuss the implications of our findings 
in the context of AGN feedback on the molecular gas. 
Finally, we summarise our conclusions in
Section~\ref{sec:conclusions}.

\section{Sample description}
\label{sec:sample}

The initial sample consists 
of 19 nearby (luminosity distance $D_L < 28$ Mpc) 
AGNs, combining the NUGA sample 
\citep[Nuclei of Galaxies,][]{combes19, audibert19, audibert21}
with the core sample of GATOS galaxies presented in
\citet{garciaburillo21}.
The GATOS core sample was selected from the
70 month \textit{Swift}-BAT all-sky 
hard-X ray survey \citep{baumgartner13}
with distances $D_{\rm L} < 28$ Mpc
and luminosities $L(14-150 \text{ keV}) \geq 10^{42}$
erg s$^{-1}$.
The CO($3-2$) data used here were analysed in detail by
\citet{garciaburillo21}, where moment maps
and kinematic modelling are discussed.
All galaxies have CO($3-2$) emission observed with the 
Atacama Large Millimeter/submillimeter Array (ALMA)
Band 7, covering a field of view of 17\arcsec{}, 
and spatial resolutions ranging
from 0.08\arcsec{} to 0.2\arcsec{},
corresponding to a physical size range of $4-16$ pc.

Inclinations and position angles of the targets are taken
from \cite{garciaburillo21}, which reported the results
of a kinematical analysis performed with the
\texttt{kinemetry} software \citep{krajnovic06}.
Since our goal is to study the clumpiness of the molecular gas,
we have decided to discard the highly inclined 
($i > 75^{\circ{}}$) galaxies from the original sample.
This decision was made to avoid severe projection effects, 
as high inclination makes it more likely
for the line of sight to intersect multiple layers 
of clumps and diffuse gas. 
This selection results in a sample of 16 galaxies, 
all with inclinations $i < 60^{\circ{}}$.
Their intrinsic $2-10$ keV luminosities 
span from $10^{39}$ to $10^{43.5}$ erg s$^{-1}$,
and their Eddington ratios, 
$\lambda_{\text{Edd}} \equiv L_{\text{AGN}} / L_{\text{Edd}}$, 
range from $-6.4$ to $-0.7$ in logarithmic scale.
Table~\ref{tab:sample} lists the basic properties 
of the galaxies analysed in this study, and
Figure~\ref{fig:sample} presents an atlas 
of their CO($3-2$) emission.

The sensitivity of the CO($3-2$) maps ranges 
from 6 to 28 mJy km s$^{-1}$ beam$^{-1}$. 
We checked whether this sensitivity correlates with 
the luminosity distance of the galaxies, and found 
no such trend (see Figure~\ref{fig:noise_jy}). 
To express these values in terms of molecular mass 
surface density, we adopted reasonable conversion factors: 
first from CO($3-2$) to CO($1-0$) ($r_{31}$), 
and then from CO($1-0$) to molecular mass via the 
CO-to-H$_2$ conversion factor ($X_{\text{CO}}$). 
We explored two extreme cases. The first assumes 
$r_{31} = 0.7$, the average value measured by 
\cite{israel20} in 126 galaxy centers, combined with 
the standard Milky Way disk value 
$X_{\text{CO}} = 2 \times 10^{20}$ 
mol cm$^{-2}$ (K km s$^{-1}$)$^{-1}$ \citep{bolatto13}.
The second adopts $r_{31} = 2.9$, 
as measured at the AGN position of NGC 1068 
by \citet{viti14}, coupled with the recommended value 
for the central region ($R < 500$ pc) of the Milky Way,
$X_{\text{CO}} = 0.5 \times 10^{20}$ 
mol cm$^{-2}$ (K km s$^{-1}$)$^{-1}$ \citep{bolatto13}. 
With the first set of parameters, the resulting 
sensitivities span $23-381$ M$_{\odot}$ pc$^{-2}$, 
while with the second they range from 1 to 23 
M$_{\odot}$ pc$^{-2}$ (see Figure~\ref{fig:noise_msun}). 
The achieved sensitivities are generally comparable 
to, or lower than, the average surface density 
of GMCs \citep[170 M$_{\odot}$ pc$^{-2}$,][]{mckee07},
and in most cases sufficient to detect individual GMCs
within a single ALMA beam at the 3$\sigma$ level.

\begin{figure*}
\centering
\includegraphics[width=0.98\textwidth]{
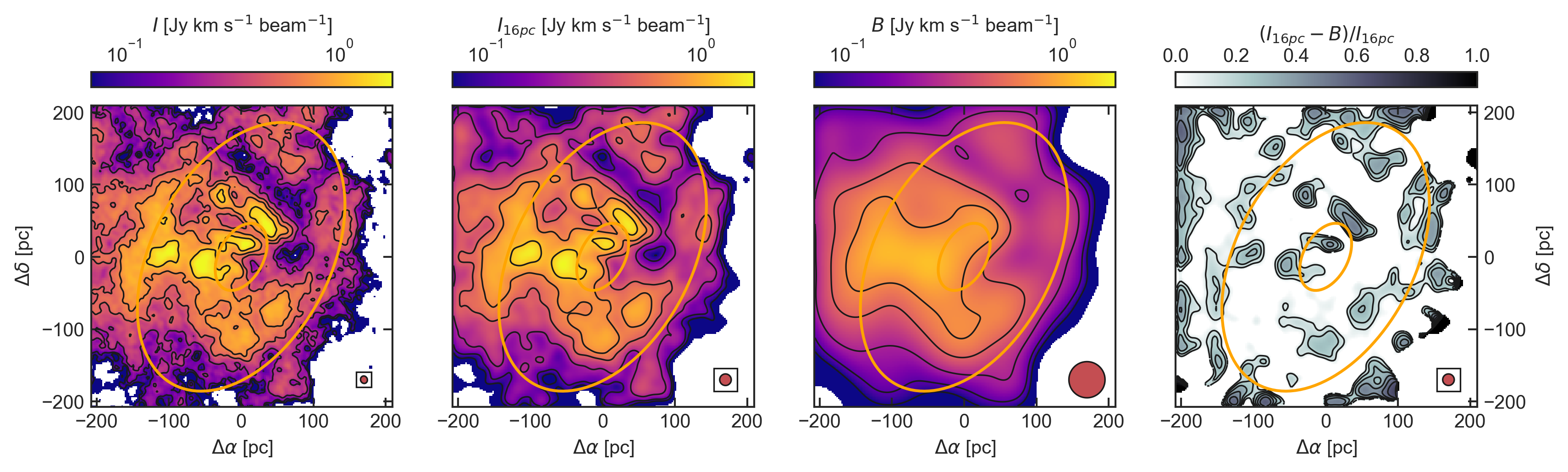}
\caption{
ALMA observations of the CO($3-2$) emission 
in the target galaxy NGC 3227 and the 
clumpiness computation applied within 
a 400 pc $\times$ 400 pc box around the AGN.
\textit{First panel}: CO($3-2$) emission 
at native resolution ($10.5 \times 9.4$ pc), 
with contours at $(3, 5, 10, 20, 30, 50) \times \sigma$.
\textit{Second panel}:
CO($3-2$) emission smoothed to a common physical
resolution of 16 pc,
using contours at the same $\sigma$ levels.
\textit{Third panel}: CO($3-2$) emission
further smoothed with a Gaussian kernel of FWHM 50 pc, 
with the same contour levels.
\textit{Fourth panel}: clumpiness map computed
with Method \#1 
(see Section~\ref{sec:methods_description}),
with contours every 0.1 from 0 to 1.
In all panels, the two ellipses correspond 
to circles of 50 and 200 pc radius in the 
galaxy plane, projected onto the sky.
The red ellipse within a square in the bottom 
right of each panel shows the native ALMA beam 
($10.5 \times 9.4$ pc in this case).
}
\label{fig:example}
\end{figure*}

\section{Data analysis}
\label{sec:analysis}

\subsection{Identifying the clumps}

To visualise and measure the clumpiness of the molecular 
gas, we chose the clumpiness parameter described by
\cite{conselice03} among those
available in the literature. 
While \cite{conselice03} originally estimated 
clumpiness (S), together with
concentration (C) and asymmetry (A),
from optical images tracing stellar light in entire galaxies
--- the so-called CAS parameters ---
this method has since been applied 
to ALMA CO($2-1$) observations by \cite{yamamoto25},
using data from the
Physics at High Angular resolution in Nearby GalaxieS
\citep[PHANGS][]{leroy21} survey.

The method involves smoothing 
the original observation and subtracting the 
smoothed map ($B$, for blurred) 
from the original intensity map ($I$). 
This results in an $I-B$ residual map, 
which highlights only the high-spatial frequency components
of the original map. 
Negative pixels (where $I-B < 0$) are set to zero. 
The clumpiness parameter ($S$) calculated by 
\cite{conselice03} is then normalised by the original 
intensity, $S = (I-B)/I$, so that it has a value between 
zero (indicating smooth emission, with no clumps)
and one (indicating high clumpiness). 
In this work, we follow the procedure outlined in
\cite{conselice03} to identify the clumps, defined as
the pixels or contiguous regions where $I-B > 0$.
We then apply three different methods to quantify 
the clumpiness of the gas, 
as described in the following sections.
Before performing any analysis,
we first smooth all the CO maps
to a common physical resolution
of 16 pc, corresponding to the largest
ALMA beam in the sample.

\subsection{Smoothing CO(3-2) maps}
\label{sec:smoothing}

A critical factor of the \cite{conselice03} procedure
is the size of the smoothing kernel. 
Ideally, we want the full width at half maximum (FWHM) 
of the kernel to be larger than the 
spatial resolution
but smaller than the aperture we are considering. 
Moreover, we expect the largest GMCs
to have sizes of $\lesssim 100$ pc
\citep[see, e.g.,][]{garcia14}, 
so smoothing over this size would result in 
blending several GMCs and diluting the high-density gas. 
Therefore, we use smoothing kernels with FWHM 
between 20
and 100 pc in the following analysis.
We will discuss the implications of different kernel sizes
in Section~\ref{sec:discussion}.

An example of the full procedure can be seen in the four 
panels of Figure~\ref{fig:example} for the galaxy
NGC 3227.
We first smooth the native-resolution 
CO($3-2$) emission map 
(\textit{first panel}) with a 16-pc Gaussian kernel 
to bring all galaxies to a common physical resolution 
(\textit{second panel}).
This map is then smoothed again with a 
larger Gaussian kernel (50 pc in Figure~\ref{fig:example}),
producing a version 
that traces the average gas distribution over the 
corresponding spatial scale (\textit{third panel}).
We subtract this smoothed map from the 16-pc map 
and retain only the pixels with positive residuals 
(\textit{fourth panel}; see details on how
clumpiness is computed below).
This procedure highlights the regions 
where the emission is significantly
more concentrated than the surrounding gas distribution,
identifying candidate clumps.

\subsection{Measuring nuclear and circumnuclear clumpiness}
\label{sec:methods_description}

In contrast to distant quasars 
\citep[see, e.g.,][]{maiolino12, genzel14, 
carniani17, bischetti19b}, 
nearby AGNs are generally found 
to primarily affect their host galaxies 
within the central kiloparsec,
mainly due to their lower luminosities 
and outflow energetics 
\citep[see, e.g.,][]{querejeta16b, fluetsch19, 
esposito22, esposito24a, esposito24b, 
garciabernete21, ramosalmeida22,
garciaburillo21, garciaburillo24}.
In this work, we follow \cite{garciaburillo21, garciaburillo24} 
in defining two concentric regions 
centred on the AGN position
\citep[based on the ALMA continuum peak,
see][]{garciaburillo21},
with radii of 50 pc and 200 pc
in the plane of the galaxy.
These scales were previously adopted to compute the 
cold molecular gas concentration index 
(CCI $\equiv \log (\Sigma_{50} / \Sigma_{200})$).
We use inclinations and position angles 
to project these two circles onto the 
plane of the galaxy.
An example of this can be seen in Figure~\ref{fig:example}, 
where the two ellipses are drawn on top
of the CO($3-2$) emission map.
We explore three different methods for measuring 
the clumpiness within the 50-pc and 200-pc apertures.

\begin{figure*}
\centering
\includegraphics[width=0.98\textwidth]{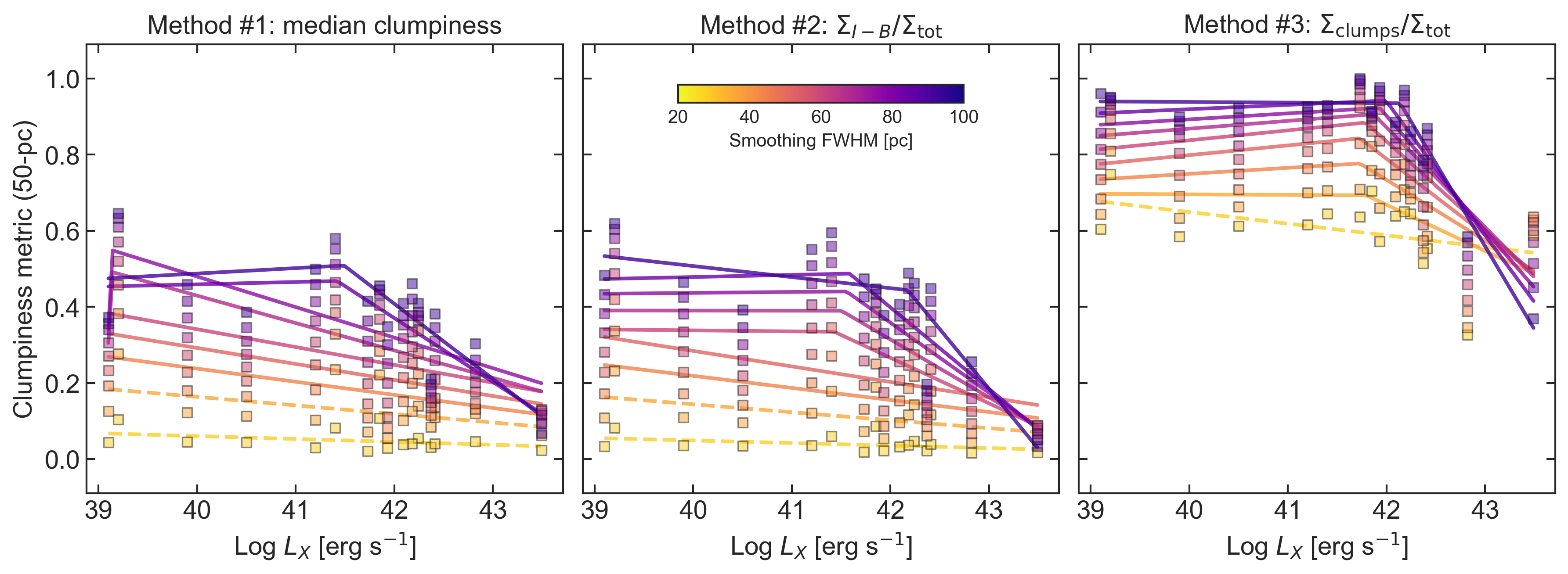}
\caption{
Clumpiness as a function
of AGN intrinsic X-ray $2-10$ keV luminosity.
Clumpiness is measured 
within the nuclear 50-pc-radius aperture
with the three different
methods described in Section~\ref{sec:methods_description}
(panels left to right).
Data points (square markers; nine per galaxy)
are coloured according to the nine smoothing 
FWHM sizes adopted
(see colour bar in the central panel).
Coloured lines show the \texttt{pwlf} fits 
to the data points, with the same colour
as the fitted dataset; lines are solid
where a linear fit is statistically accepted, 
and dashed where it is rejected 
(see Section~\ref{sec:stat_methods}).
}
\label{fig:smooths_comparison}
\end{figure*}

\subsubsection{Method \#1: median clumpiness}
\label{sec:method1}

The first method follows that of \cite{conselice03}: 
we calculate the clumpiness as $(I-B)/I$ 
pixel by pixel, setting negative pixels to zero, 
as shown in the fourth panel of Figure~\ref{fig:example}. 
We then take the median value of the clumpiness 
within the two apertures, referring to this method 
as "Method \#1: median clumpiness". 
To estimate the errors, we use the $1\sigma$ distance 
from the median values. 
This method provides a clumpiness value that can be
compared with other studies
\citep[e.g.][]{conselice03, yamamoto25}. 
However, high clumpiness values tend to identify 
regions with higher contrast, especially 
in the outskirts of the CO emission, 
rather than dense or massive clumps, 
as visible in Figure~\ref{fig:example}. 
For this reason, we sought alternative methods 
to estimate the fraction of gas in clumps.

\subsubsection{Method \#2: 
\texorpdfstring{$\Sigma_{I-B}/\Sigma_{\text{tot}}$}{Sigma(I-B)/Sigma(tot)}}
\label{sec:method2}

The second method is similar to Method \#1 in principle, 
as it calculates the clumpiness as $(I-B)/I$. 
However, instead of computing it pixel by pixel
and then taking the median value, we separately 
sum the $I-B$ and $I$ values within the 
apertures and then divide them. 
We call this method 
"Method \#2: $\Sigma_{I-B}/\Sigma_{\text{tot}}$".
Since CO intensity can be converted to molecular mass 
using a CO-to-H$_2$ conversion factor $\alpha_{CO}$, 
and assuming the same $\alpha_{CO}$ for both 
the clumpy emission and the underlying smooth emission, 
the result of $\Sigma_{I-B}/\Sigma_{\text{tot}}$ can be interpreted 
as the mass fraction of gas in clumps, 
independent of the choice of $\alpha_{CO}$. 
This method returns values between 0 and 1, 
where higher values indicate a greater contrast between clumpy
and smooth gas. 
Since this method is analogous to aperture photometry, 
we use the photometric errors to estimate 
the uncertainties. For the ALMA observations, 
these are dominated by the calibration error, 
which can be as high as $\sim 10\%$
\citep{francis20}. 
This results in a propagated uncertainty of $\sim 17\%$.

\subsubsection{Method \#3: 
\texorpdfstring{$\Sigma_{\text{clumps}} / \Sigma_{\text{tot}}$}{
Sigma(clumps)/Sigma(tot)}}
\label{sec:method3}

The third method uses the condition $I-B > 0$ 
to identify the clumps and sums the original pixel 
intensities $I$ (without subtracting the smoothed $B$ emission) 
to calculate the flux of the clumps, $\Sigma_{\text{clumps}}$, 
within the different apertures. 
We then divide this by the total flux $\Sigma_{\text{tot}}$, 
and refer to this as 
"Method \#3: $\Sigma_{\text{clumps}} / \Sigma_{\text{tot}}$".
By not subtracting the smoothed emission, we aim to obtain 
a more direct measurement of the fraction of molecular 
mass in clumps. 
As a result, this method tends to yield systematically 
higher clumpiness values compared to the other two.
As with the second method, we assume a 
$\sim 10\%$ error on the flux measurements, 
which propagates to a $\sim 14\%$ uncertainty.

These three methods are designed to capture 
complementary aspects of clumpiness, 
each with its own sources of systematic uncertainty. 
Using multiple approaches allows us to test the 
robustness of our results against these effects, 
rather than relying on a single estimator of clumpiness.

\subsection{Estimating the influence of AGN on clumpiness}
\label{sec:stat_methods}

One of the most reliable indicators of the power 
of an AGN is its intrinsic $2-10$ keV
X-ray luminosity,
$L_{\text{X}}$, which serves 
as a proxy for both radiative feedback on the ISM 
\citep[through the creation of X-ray dominated regions, 
or XDRs; see, e.g.,][]{maloney96, wolfire22, 
esposito24b, tadaki25}
and mechanical feedback 
\citep[through the interaction of AGN winds with 
the host galaxy's ISM; see, e.g.,][]{
fauchergiguere12, cicone14, nims15, fiore17, esposito24a}.
We use $L_{\text{X}}$ as a measure of AGN power 
and investigate whether there are correlations between
the AGN emission and our 
different clumpiness measurements, 
considering various smoothing kernels and aperture sizes. 
To do this, we apply the Piecewise Linear Fitting (PWLF) 
algorithm using the Python package \texttt{pwlf} 
\citep{pwlf}, which fits continuous piecewise linear functions
to the data based on a specified number of line segments.
We begin by fitting both a single-segment 
(i.e., standard linear regression) 
and a two-segment model. We then perform an 
F-test (at the 95\% confidence level) 
to evaluate whether the two-segment model 
provides a statistically significant improvement 
over the single-segment model. 
If the two-segment fit is not preferred, 
we adopt the single-segment model 
and assess the significance of its slope 
with a two-tailed \textit{t}-test under 
the null hypothesis of zero slope.

For each dataset (i.e., a combination of clumpiness method, 
smoothing level, and aperture) 
where a two-segment fit is preferred, 
we test for the presence of a breakpoint
--- an $L_{\text{X}}$ value where the slope changes ---
using two statistical techniques.
First, we perform Leave-One-Out Cross-Validation (LOOCV), 
iteratively excluding one data point at a time 
to examine the distribution of breakpoints within the dataset. 
This helps assess whether the fit is significantly 
influenced by individual points. 
We consider a breakpoint reliable only if 
the standard deviation
of its LOOCV distribution
is smaller than 10\% 
of the $L_{\text{X}}$-range of the data.
Second, for each validated dataset,
we run 1000 Monte Carlo simulations, 
taking into account 
the data uncertainties in both axes:
we assume a $\pm0.15$ dex uncertainty for $L_{\text{X}}$
\citep[following][]{garciaburillo24}, 
while clumpiness uncertainties are described
in Section~\ref{sec:methods_description}.
This approach allows us to statistically confirm 
the presence of a breakpoint and to estimate 
its $1\sigma$ confidence interval 
using Monte Carlo resampling.

\begin{figure*}
\centering
\includegraphics[width=0.98\textwidth]{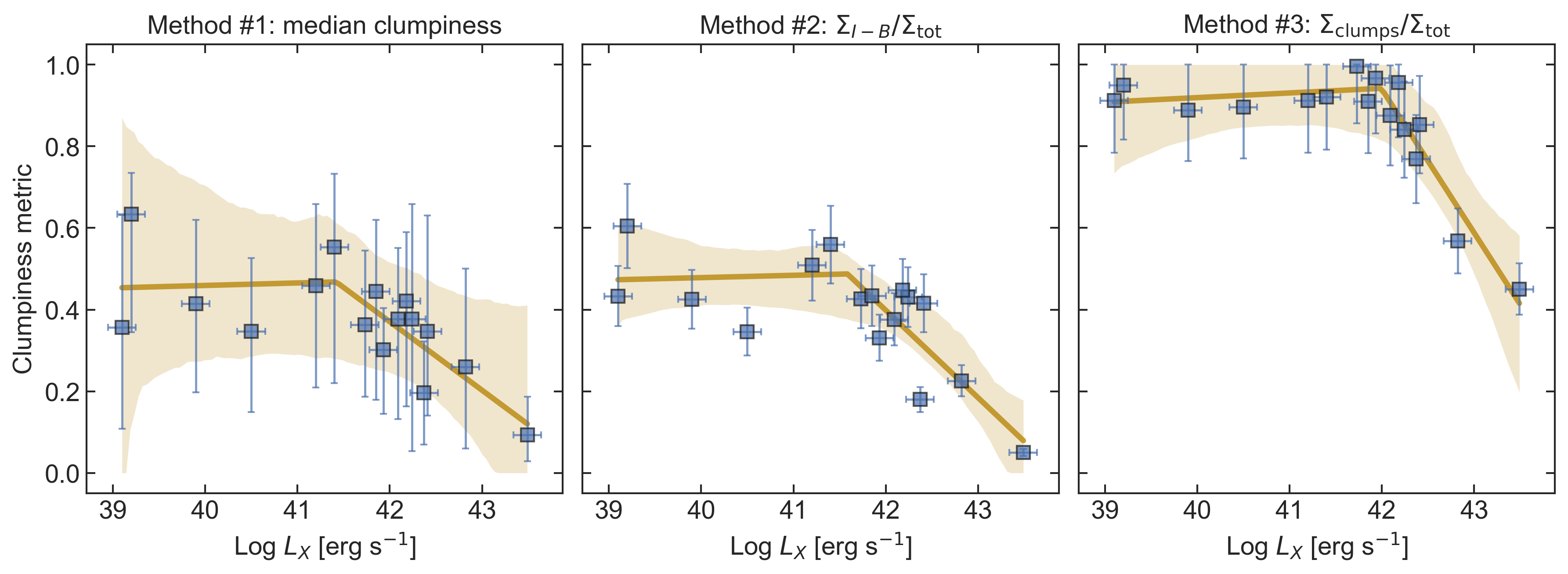}
\caption{
Clumpiness as a function
of AGN X-ray luminosity.
Clumpiness is measured 
within the nuclear 50-pc-radius aperture
with the three different
methods described in Section~\ref{sec:methods_description}
(panels left to right).
Data points are shown as blue square markers 
with error bars, computed for a 
smoothing kernel FWHM of 90 pc.
The broken orange lines indicate the 
\texttt{pwlf} fits to the data, 
and the shaded regions show the corresponding 
95\% confidence intervals.
}
\label{fig:80pc_comparison}
\end{figure*}

\section{Results}
\label{sec:results}

\subsection{The nuclear region}

Figure~\ref{fig:smooths_comparison} shows the clumpiness, 
measured within the 50-pc-radius aperture,
as a function of the X-ray luminosity, $L_{\text{X}}$, 
of our sample targets.
Clumpiness is computed using the three methods 
described in Section~\ref{sec:methods_description}, 
applying nine different smoothing levels, 
ranging from 20 to 100 pc in steps of 10 pc. 
This allows us to trace how clumpiness varies 
as a function of the smoothing scale.

For each smoothing level, we fit the data using 
the PWLF algorithm and plot the corresponding regression lines. 
At low smoothing levels, when the kernel size is close 
to the ALMA beam, the clump identification condition 
($I-B>0$) primarily selects the peaks of the clumps. 
As a result, Methods \#1 and \#2 return values close 
to zero with no apparent breakpoints in the fit.
In contrast, Method \#3 starts from a baseline value 
of $\Sigma_{\text{clumps}} / \Sigma_{\text{tot}} \approx 0.6$, 
indicating that even at the smallest smoothing scales, 
clumps already contribute approximately 60\% 
of the CO emission.
In addition, Method \#3 already shows a 
statistically significant breakpoint at
$L_{\text{X}} \sim 10^{41.8}$ erg s$^{-1}$
emerging from smoothing scales as small as 30 pc.

Up to a smoothing scale of 50 pc, the data show a 
significant ($p < 0.05$), weak ($r^2 < 0.4$)
anti-correlation between $L_{\text{X}}$ and clumpiness
for Methods \#1 and \#2.
As the smoothing level increases, 
a two-segment trend emerges
also for these methods, 
with a relatively flat trend at the lowest luminosities,
followed by a steep decline.
The breakpoint in these cases typically falls around
$L_{\text{X}} \sim 10^{41.5}$ erg s$^{-1}$. 
The exact smoothing level required to reveal 
a significant breakpoint depends on the method used: 
for Method \#3, 
the trend appears at 30 pc;
for Method \#2, at 60 pc;
and for Method \#1, at 90 pc.

From a smoothing level of 90 pc, 
a two-segment fit appears for every method. 
Figure~\ref{fig:80pc_comparison} shows the data points 
for this smoothing level, including error bars, 
the \texttt{pwlf} lines, and the 95\% confidence interval 
from Monte Carlo simulations. For Method \#1, 
the LOOCV algorithm finds a high standard deviation 
($30\%$ of the $L_{\text{X}}$ range),
indicating that the exact position of the 
$L_{\text{X}}$ breakpoint is not well constrained 
and should be treated with caution. 
This is further illustrated in 
Figure~\ref{fig:smooths_comparison}, 
where at smoothing scales of 70 and 80 pc, 
the breakpoint appears at very low 
$L_{\text{X}} \sim 10^{39}$ erg s$^{-1}$
--- probably an artifact of the fitting process ---
before shifting to higher $L_{\text{X}}$ 
values at larger smoothing scales.
The broad confidence interval 
in Figure~\ref{fig:80pc_comparison} further reflects this
uncertainty. In contrast, for Methods \#2 and \#3, 
the LOOCV validates the breakpoints, 
with standard deviations of 
7\% and 1\%
of the $L_{\text{X}}$ range, 
respectively, yielding 
$\log (L_{\text{X}} / \mbox{erg s}^{-1})_{\text{break}} = 
41.99_{-0.62}^{+0.44}$ and
$42.03_{-0.34}^{+0.32}$.

It is interesting to note that, 
for both Method \#2 and \#3,
the breakpoint position shifts to higher X-ray luminosity 
values as we increase the smoothing:
at 80-pc smoothing, we find the breakpoints at
$\log (L_{\text{X}} / \text{erg s}^{-1})_{\text{break}} = 
41.76_{-0.45}^{+0.60}$
and $41.92_{-0.36}^{+0.30}$, respectively 
for Methods \#2 and \#3,
while at 100-pc
smoothing we find them at
$\log (L_{\text{X}} / \text{erg s}^{-1}) = 
42.04_{-0.62}^{+0.49}$
and $42.12_{-0.33}^{+0.32}$.
Considering only Methods \#2 and \#3,
we find that the breakpoints are compatible with
each other, given the uncertainties, to a value of 
$L_{\text{X}} \approx 10^{42}$ erg s$^{-1}$.

Finally, we explored the relation between 
clumpiness and the AGN Eddington ratio, 
$\lambda_{\text{Edd}}$, 
as shown in Figure~\ref{fig:Edd_50pc}.
We find a similar pattern (flat regime
followed by a steep decline) by
using Methods \#1 and \#3,
with breakpoints between 
$\lambda_{\text{Edd}} = -3$ and $-2$.

\subsection{The circumnuclear region}

\begin{figure*}
\centering
\includegraphics[width=0.98\textwidth]{
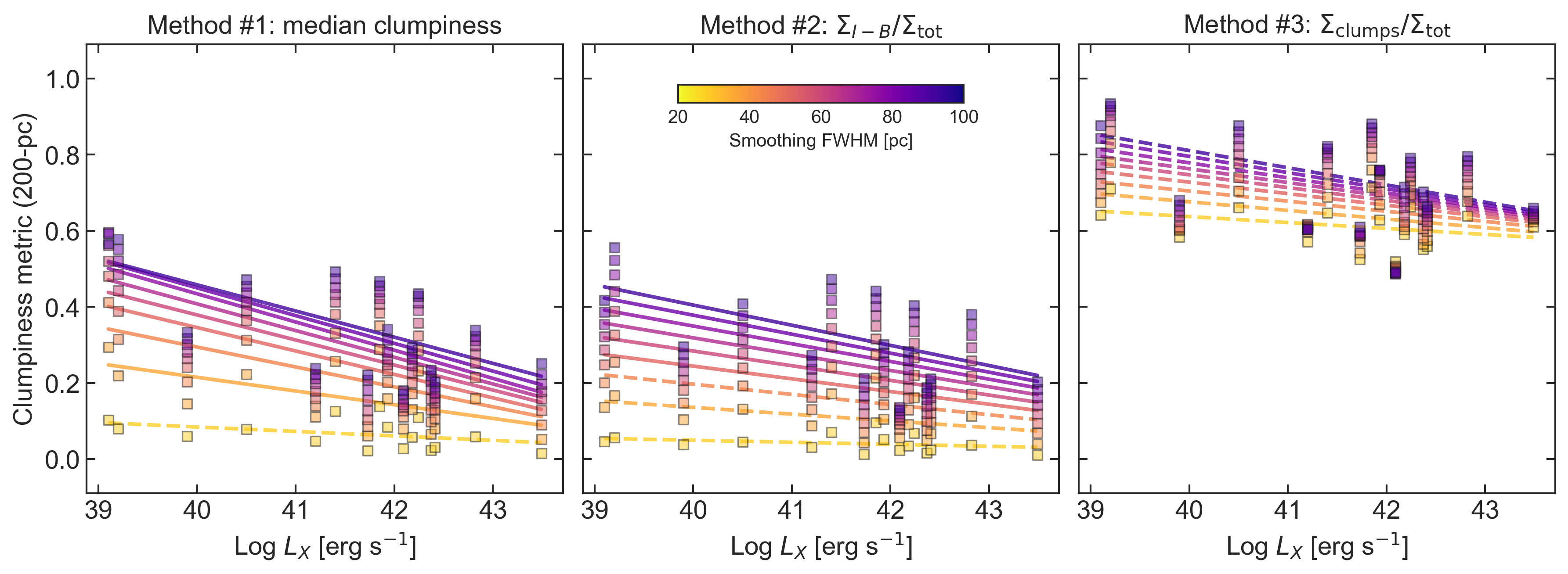}
\caption{
Clumpiness measured within the larger 200-pc-radius
aperture, plotted as a function of AGN X-ray luminosity,
Panels, lines, and markers are as in
Figure~\ref{fig:smooths_comparison}.
}
\label{fig:200pc_smooths}
\end{figure*}

Applying the same methods to the 200-pc aperture, 
we find significantly decreasing linear trends 
without breakpoints for Methods \#1 and \#2. 
Method \#3 also shows an anticorrelation 
between $L_{\text{X}}$ and clumpiness, but no smoothing kernel 
size yields a significance above $2\sigma$
(see Figure~\ref{fig:200pc_smooths}). 
Beyond physical explanations (discussed in 
Section~\ref{sec:discussion}), 
some targets in this aperture include regions dominated 
by noise or empty space (e.g., NGC 7213 and NGC 1365; 
see Figure~\ref{fig:sample}),
which may affect the analysis. 
When considering only the annular region between 
galactocentric distances of 50 pc and 200 pc, 
we see less significant results,
probably for the high noise present
in this annular aperture
(see Figure~\ref{fig:annular_smooths}).

For $L_{\text{X}} > 10^{41.5}$ erg s$^{-1}$, local 
AGNs exhibit an anticorrelation between $L_{\text{X}}$ 
and the molecular gas concentration index, 
defined as the ratio of surface densities 
within the 50-pc and 200-pc apertures 
\citep[see][]{garciaburillo24}. 
To compare with these findings, 
we calculate the ratio of clumpiness estimates 
within the two apertures for each method and smoothing level. 
As shown in Figure~\ref{fig:RoR200pc_smooths},
with all the three methods we observe a rising trend, 
rather than a flat regime, followed by a decline
in the clumpiness ratio as a function of $L_{\text{X}}$,
with a break again around $10^{42}$ erg s$^{-1}$.
However, substantial scatter is present, likely due
to the noise in the larger 200-pc aperture.

\subsection{Clumpiness vs. concentration}

In Figure~\ref{fig:50pc_vs_CCI} we explore 
the relation between the clumpiness measured 
within the central 50-pc aperture and the 
cold molecular gas concentration index 
(CCI $\equiv \log (\Sigma_{50} / \Sigma_{200})$). 
The CCI provides a more global measure of how 
centrally concentrated the cold molecular gas is, 
while the clumpiness values in 
Figure~\ref{fig:50pc_vs_CCI}
refer to the nuclear region alone.

With all three methods we observe 
a rising trend of clumpiness 
with increasing CCI up to 
CCI $\sim 0.1$, after which the relation flattens. 
This breakpoint value is significantly lower 
than the one reported by \cite{garciaburillo24} 
in the CCI–$L_{\text{X}}$ relation, 
where the transition occurs around CCI $\sim 0.5$.

In Figure~\ref{fig:RoR200pc_CCI} we present the 
plot of the clumpiness ratio 
(measured within 50 pc divided by that within 
200 pc) as a function of the CCI.
In this case, all three methods reveal 
significant correlations, with a clear break 
at CCI $\sim 0.6$. This breakpoint is closer 
to the value found by \cite{garciaburillo24} 
in the CCI–$L_{\text{X}}$ relation.

\section{Discussion}
\label{sec:discussion}

\subsection{Comparison with previous studies}

Method \#1 allow us to compare our
measured clumpiness values with existing literature,
as it is the same method originally developed
by \cite{conselice03}.
In their work, median clumpiness, measured
from stellar light in the optical R-band, 
ranges from 0 for elliptical galaxies to $\sim 0.7$
for starburst galaxies, with data 
smoothed using a kernel of size $0.3 \times R(\eta=0.2)$,
where $R(\eta)$ is the Petrosian
radius\footnote{The Petrosian radius is defined as 
the location where the ratio of surface brightness at a radius, 
divided by the surface brightness within the radius,
reaches some value $\eta$ \citep{petrosian76}.}.
The median CO($2-1$) clumpiness in a subsample of PHANGS,
mostly composed of spiral galaxies,
is $0.27 \pm 0.08$ \citep{yamamoto25},
where the ALMA data (at 180 pc resolution)
were smoothed
with a kernel of size $0.3 \times R_e$, where 
the effective radius $R_e$ of the galaxy
was chosen rather than $R(\eta=0.2)$ to better
represent the spatial extent of CO($2-1$):
for example, for the PHANGS galaxy NGC 3627,
$R_e \approx 66$\arcsec{} while
$R(\eta=0.2) \approx 178$\arcsec{}
(3.6 and 9.7 kpc, respectively).

In our case, using Method \#1, we find the median
clumpiness in the nuclear 50-pc aperture to range from
$0.08_{-0.03}^{+0.03}$ to $0.39_{-0.10}^{+0.11}$
(with the upper and lower values representing the $1\sigma$
deviation of the sample), depending on the smoothing
kernel size ($20-100$ pc,
see Figure~\ref{fig:smooths_comparison}).
In the larger 200-pc aperture, these values become
$0.08^{+0.07}_{-0.04}$ and $0.34^{+0.13}_{-0.12}$,
respectively for 20-pc and 100-pc smoothing
(see Figure~\ref{fig:200pc_smooths}).

These results may at first suggest that 
the molecular gas clumpiness in the nuclear regions 
of AGN-host galaxies is comparable to that observed 
in the discs of normal spirals (PHANGS) 
and much lower than the values found for 
the stellar emission of starburst galaxies.
However, a direct comparison is not straightforward, 
because the measured clumpiness depends critically 
on the spatial scale traced by the data and 
on the chosen smoothing kernel.
In our case, we probe only the central few hundred parsecs 
at a much higher resolution (16 pc), 
and by applying different smoothing kernels 
(from 20 to 100 pc) we can highlight structures 
of different physical sizes.
In contrast, the study by \cite{yamamoto25}
covers entire galaxy discs with lower resolution 
(180 pc), and their smoothing kernel is 
several kpc, making it more sensitive 
to large-scale molecular complexes.

\begin{figure*}
\centering
\includegraphics[width=0.98\textwidth]{
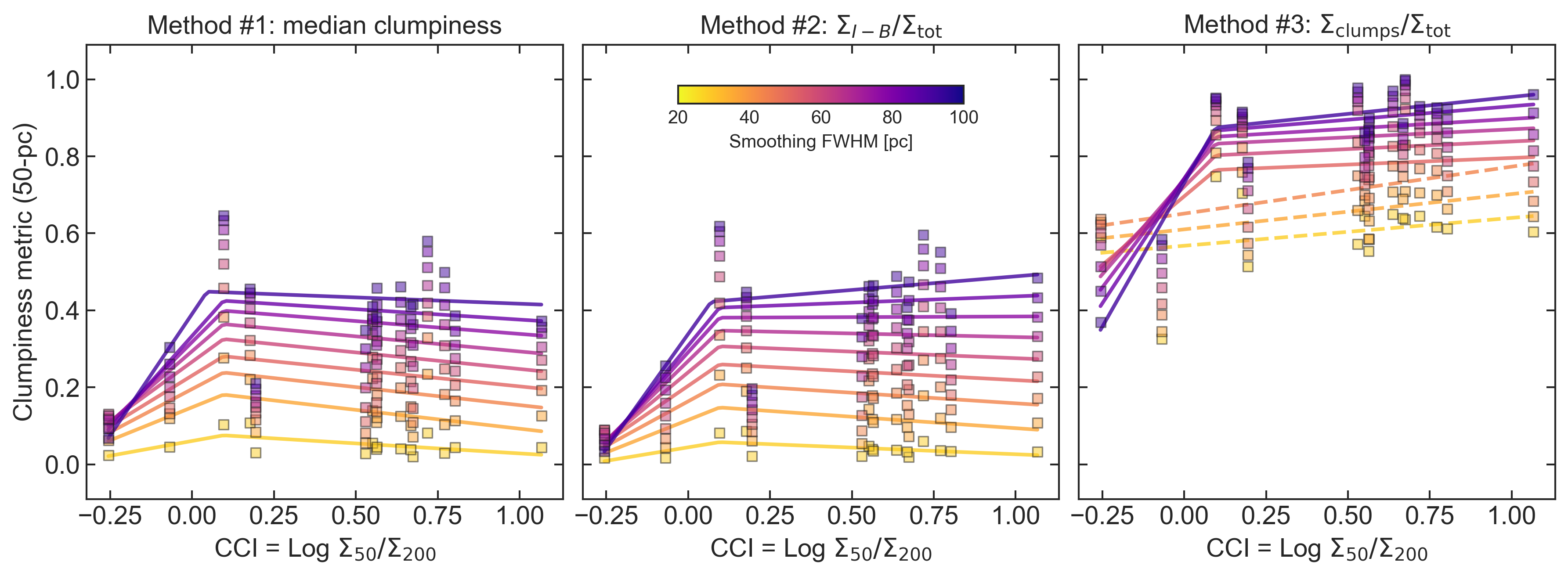}
\caption{
Clumpiness measured 
within the 50-pc-radius aperture, 
plotted as a function of cold molecular gas 
concentration index (CCI 
$\equiv \log (\Sigma_{50} / \Sigma_{200})$),
where $\Sigma_r$ is the surface
density within an aperture
of radius $r$ parsec.
Panels, lines, and markers are as in
Figure~\ref{fig:smooths_comparison}.
}
\label{fig:50pc_vs_CCI}
\end{figure*}

\subsection{Clumpiness as a function of X-ray luminosity}

By exploring clumpiness as a function of smoothing scale, 
we find distinct behaviours depending on the 
method adopted. With Methods \#1 and \#2,
small-scale structures 
(highlighted by smoothing kernels $\lesssim 50$ pc) 
show generally low clumpiness values 
and an anti-correlation with AGN luminosity. 
At larger smoothing scales ($\gtrsim 80$ pc), 
clumpiness remains roughly constant with AGN luminosity 
up to $L_{\text{X}} \approx 10^{42}$ erg s$^{-1}$, 
followed by a clear drop at higher luminosities. 
This behaviour may suggest that larger molecular structures 
($\sim 80-100$ pc) are more resilient to 
moderate AGN feedback (as traced by $L_{\text{X}}$), 
but are eventually disrupted 
when the AGN becomes sufficiently powerful.

In contrast, Method \#3 consistently reveals 
a statistically significant breakpoint around 
$L_{\text{X}} \approx 10^{42}$ erg s$^{-1}$, 
even at the smallest scales ($\sim 30$ pc).
This indicates that, regardless of the physical scale 
probed, the fraction of emission associated with 
compact clumps decreases sharply once the AGN reaches 
high luminosities. Such behaviour supports the idea 
that AGN activity can impact both small ($\lesssim 50$ pc) 
and large ($\gtrsim 80$ pc) molecular structures, 
potentially through a combination of X-ray irradiation, 
winds, and cosmic rays.

The different emergence of the breakpoint between 
Methods \#2 and \#3 likely reflects the 
stronger sensitivity of Method \#2 to the contrast 
between clumps and background. 
At small smoothing scales, more luminous AGNs seem 
to reduce this contrast, leading to lower values of
$\Sigma_{I-B}$. Method \#3, which does not subtract 
the background, is less affected and therefore
reveals the breakpoint already at small scales. 
At larger smoothing ($\geq 60$ pc),
the background level decreases and
the contrast naturally diminishes, 
causing the two methods to converge.

The observed flat regime
followed by a decline in clumpiness at 
$L_{\text{X}} \approx 10^{42}$ erg s$^{-1}$ ,
could be the result of a
balance between significant gas inflow 
from the outer regions
\citep[driven by gravitational torque, see e.g.][]{
garciaburillo05, casasola08, combes14, 
querejeta16, audibert19, audibert21}
and the mild AGN feedback acting on molecular gas 
in galaxies below the luminosity breakpoint.
In this scenario, the nuclear regions 
(within 50 pc from the AGN) 
would maintain their clumpiness up to a certain AGN 
luminosity thanks to the continuous feeding 
of fresh molecular material, until feedback becomes 
strong enough to dominate and disperse 
also the larger molecular complexes.

We note that the three methods we developed to
quantify clumpiness, despite being sensitive 
to different systematics, consistently reveal 
the same negative trend with $L_{\text{X}}$.
This agreement strengthens the robustness 
of our finding that higher AGN luminosity 
is associated with smoother gas distributions.

Interestingly, when measuring clumpiness in the 
larger aperture (200 pc radius,
Figure~\ref{fig:200pc_smooths}), we do not observe 
the same turnover: instead, we find a continuous 
anti-correlation with AGN luminosity, 
independently of the smoothing kernel size.
This may be because, in the larger aperture, 
we lose the direct view of the nuclear clumpiness 
and instead average over a broader region of the 
circumnuclear disk.
In this larger region, the inflow of molecular gas 
at low $L_{\text{X}}$ may be not sufficient to balance 
the AGN feedback at any of the smoothing scales considered,
resulting in a continuous anti-correlation 
between clumpiness and AGN luminosity.
It is also worth noting that, in this larger aperture,
the results for some galaxies may be affected 
by higher noise levels,
which can bias the clumpiness measurement.
Furthermore, when we isolate the annular region
between 50 and 200 pc to exclude the very nucleus,
the anti-correlation with AGN luminosity 
nearly disappears (see Figure~\ref{fig:annular_smooths}), 
suggesting that the observed trend
is mostly driven by the nuclear region itself.

The clumpiness measured in the central 50 pc 
could in principle reflect the efficiency of AGN feeding:
our analysis 
indeed shows a significant correlation
(especially with Methods \#1 and \#3)
with the AGN Eddington ratio
$\lambda_{\text{Edd}}$ (Figure~\ref{fig:Edd_50pc}).
This may suggest that the small-scale molecular 
gas structure is impacted not only
by the AGN radiative power traced by $L_{\text{X}}$,
that is, by the number of high-energy photons,
but also by the shape of the spectral energy
distribution and/or the black hole mass,
both of which influence $\lambda_{\text{Edd}}$.

\subsection{Clumpiness as a function of gas concentration}

Our analysis of the molecular gas clumpiness 
within the central 50 pc shows
complex behaviour when compared to the 
cold molecular gas concentration index 
(CCI), which quantifies the radial concentration
of molecular gas between 50 and 200 pc scales
(see Figure~\ref{fig:50pc_vs_CCI}).
At any smoothing scale,
clumpiness seems to be correlated with CCI, 
with a breakpoint around CCI $\sim 0.1$.
The absence of a 1:1 correlation imply 
that other hidden variables, likely including AGN 
luminosity and feedback processes, also 
influence gas morphology. 
Notably, in the clumpiness ratio vs. CCI 
plot presented in Figure~\ref{fig:RoR200pc_CCI}, 
a similar correlation emerges with a breakpoint 
around CCI $\sim 0.6$, close to the break found 
in the CCI vs. $L_{\text{X}}$ plot
by \cite{garciaburillo24}.

We note that a correlation between
clumpiness and concentration was also reported by
\cite{conselice03} for stellar light emission
on kpc scales: in their case, clumpiness
correlated with H$\alpha$ emission from
young stars, while concentration was linked to
the bulge-to-total light ratio.
In our study, these quantities are measured
from molecular gas emission within the nuclear 400 pc.
Nonetheless,
the observed structural change in the molecular gas
--- from a clumpier to a smoother distribution 
with increasing AGN luminosity --- may have important
implications for the host galaxy.
A smoother gas distribution could indicate
the suppression of dense molecular clouds, 
thereby reducing the local star formation rate, 
and at the same time make the gas more vulnerable 
to being entrained and expelled by AGN-driven winds 
\citep[see, e.g.,][]{ward24}.
At lower $L_{\text{X}}$, on the other hand,
the higher degree of
clumpiness may negatively affect the propagation of jets 
and winds, causing them to scatter and dissipate 
their kinetic energy more efficiently into the 
surrounding ISM 
\citep{perucho12, wagner12, mukherjee18, tanner22}.

\section{Conclusions}
\label{sec:conclusions}

In this paper, we investigate several methods
for measuring the clumpiness of the cold molecular gas
in the nuclear region ($50-200$ pc) of galaxies,
and its relationship with AGN activity, as measured
by the X-ray $2-10$ keV luminosity $L_{\text{X}}$,
and with cold molecular gas concentration,
measured by the index
CCI $\equiv \log (\Sigma_{50} / \Sigma_{200})$.

By analysing how clumpiness changes with smoothing scale, 
we find that small-scale molecular structures 
(revealed by smoothing $\lesssim 50$ pc) 
generally show lower clumpiness values and an 
anti-correlation with AGN luminosity across 
the entire range of $L_{\text{X}}$ considered
(see Figure~\ref{fig:smooths_comparison}).
In contrast,
when probing larger molecular structures 
($\sim 70-100$ pc)
---
or when estimating clumpiness more directly, 
by summing over the clumps without subtracting 
the underlying background emission ---
we observe a different behaviour: 
clumpiness remains roughly constant up to a threshold
luminosity of $L_{\text{X}} \approx 10^{42}$ erg s$^{-1}$,
beyond which it drops sharply
(see Figure~\ref{fig:80pc_comparison}).
This suggests that larger gas clumps may resist 
AGN feedback at moderate luminosities, 
possibly due to ongoing inflow that replenishes 
the nuclear region, 
but are eventually disrupted at higher AGN power.
When measuring clumpiness in the larger 400-pc aperture, 
however, this turnover effect is not observed: 
we instead find a continuous anti-correlation 
with $L_{\text{X}}$.
We also find evidence for a correlation between 
clumpiness and concentration (CCI), 
particularly at low CCI values (CCI $< 0$), 
while at higher CCI the relation becomes weaker 
and consistent with a plateau
(see Figure~\ref{fig:50pc_vs_CCI}).

Our interpretation is that AGN feedback partially
destroys nuclear molecular gas
\citep[hence decreasing its concentration;][]{
garciaburillo21, garciaburillo24},
and smooths the surviving clouds,
resulting in a less clumpy medium.
We interpret this as evidence of negative AGN feedback 
in the immediate surroundings of the AGN, 
where the gas becomes less capable 
of aggregating into dense clouds ---
a necessary condition for star formation.

It is worth noting that these results were made possible
by the high spatial resolution (16 pc)
of the gas observations,
which allow us to probe the molecular gas in 
unprecedented detail.
In contrast, hydrodynamical simulations still face challenges
in capturing the full complexity of the molecular ISM,
particularly in modelling the interaction between
AGN winds and a clumpy medium, 
due to the coexistence of vastly different
physical scales and gas phases
\citep[e.g.][]{mccourt18, meenakshi22, ward24}.

The results presented here not only reveal a negative 
correlation between clumpiness and AGN activity, 
but also highlight the distinct impact of AGN feedback
on gas structures at small spatial scales, 
an aspect that has been poorly investigated 
in previous studies.
As clumpiness becomes an increasingly 
important parameter for characterising both nearby
\citep{leroy13b, sun22, yamamoto25}
and distant galaxies
\citep{murata14, sok25},
our analysis offers a complementary perspective 
on how AGN activity may shape the 
internal structure of the molecular gas.

\begin{acknowledgements}
We thank the anonymous referee for suggestions and comments 
that helped to improve the presentation of this work.
We acknowledge the use of Python \citep{python3}
and the following libraries:
Astropy \citep{astropy1, astropy2},
Matplotlib \citep{matplotlib},
NumPy \citep{numpy},
Pandas \citep{pandas},
Photutils \citep{photutils},
pwlf \citep{pwlf},
Seaborn \citep{seaborn}, and
SciPy \citep{scipy}.
This paper makes use of the following ALMA data:
ADS/JAO.ALMA\#2015.0.00404.S,
\#2016.1.00254.S,
\#2016.0.00296.S,
\#2017.1.00082.S, 
and \#2018.1.00113.S.
ALMA is a partnership 
of ESO (representing its member states), NSF (USA) 
and NINS (Japan), together with NRC (Canada), MOST 
and ASIAA (Taiwan), and KASI (Republic of Korea), 
in cooperation with the Republic of Chile. 
The Joint ALMA Observatory is operated by ESO, 
AUI/NRAO and NAOJ.
This research has made use of the 
NASA/IPAC Extragalactic Database (NED), 
which is funded by the National Aeronautics 
and Space Administration 
and operated by the 
California Institute of Technology.
FE acknowledges support from ESA through the 
Science Faculty - Funding reference ESA-SCI-E-LE-092.
FE, SGB, MQ, and AU acknowledge support from the Spanish grant
PID2022-138560NB-I00, funded by
MCIN/AEI/10.13039/501100011033/FEDER, EU.
AAH acknowledges support from grant 
PID2021-124665NB-I00
funded by MCIN/AEI/10.13039/501100011033 
and by ERDF A way of making Europe.
AA acknowledges funding from the European Union 
grant WIDERA ExGal-Twin, GA 101158446.
AJB acknowledges funding from the “FirstGalaxies” 
Advanced Grant from the European Research Council (ERC) 
under the European Union’s Horizon 2020 research 
and innovation program (Grant agreement No. 789056).
CRA and AA acknowledge support from project 
“Tracking active galactic nuclei feedback from 
parsec to kiloparsec scales”, 
with reference PID2022-141105NB-I00.
EB acknowledges support from the Spanish grants 
PID2022-138621NB-I00 and PID2021-123417OB-I00, 
funded by MCIN/AEI/10.13039/501100011033/FEDER, EU.
IGB is supported by the Programa Atraccíon 
de Talento Investigador "César Nombela" via grant 
2023-T1/TEC-29030 funded by the Community of Madrid.
OGM acknowledge financial support for PAPIIT/UNAM project 
IN1097123 and Ciencia de Frontera (SECIHTI) project CF-2023-G100.
SFH acknowledges support through UK Research and Innovation 
(UKRI) under the UK government’s Horizon Europe 
funding Guarantee 
(EP/Z533920/1) and an STFC Small Award (ST/Y001656/1).
\end{acknowledgements}

\bibliographystyle{aa} 
\bibliography{bibespo}

\begin{appendix}

\section{Ancillary material: sensitivities, emission maps, 
clumpiness analysis, and additional correlations}

In this appendix we present the main ancillary 
information and visual material used in this study. 
Table~\ref{tab:sample} lists the basic properties 
of the 16 galaxies analysed: 
equatorial coordinates, luminosity distances, 
hard X-ray luminosity ($L_{\text{X}}$), 
Eddington ratios ($\lambda_{\text{Edd}}$),
inclinations, and position angles.

\begin{figure}
\centering
\includegraphics[width=0.49\textwidth]{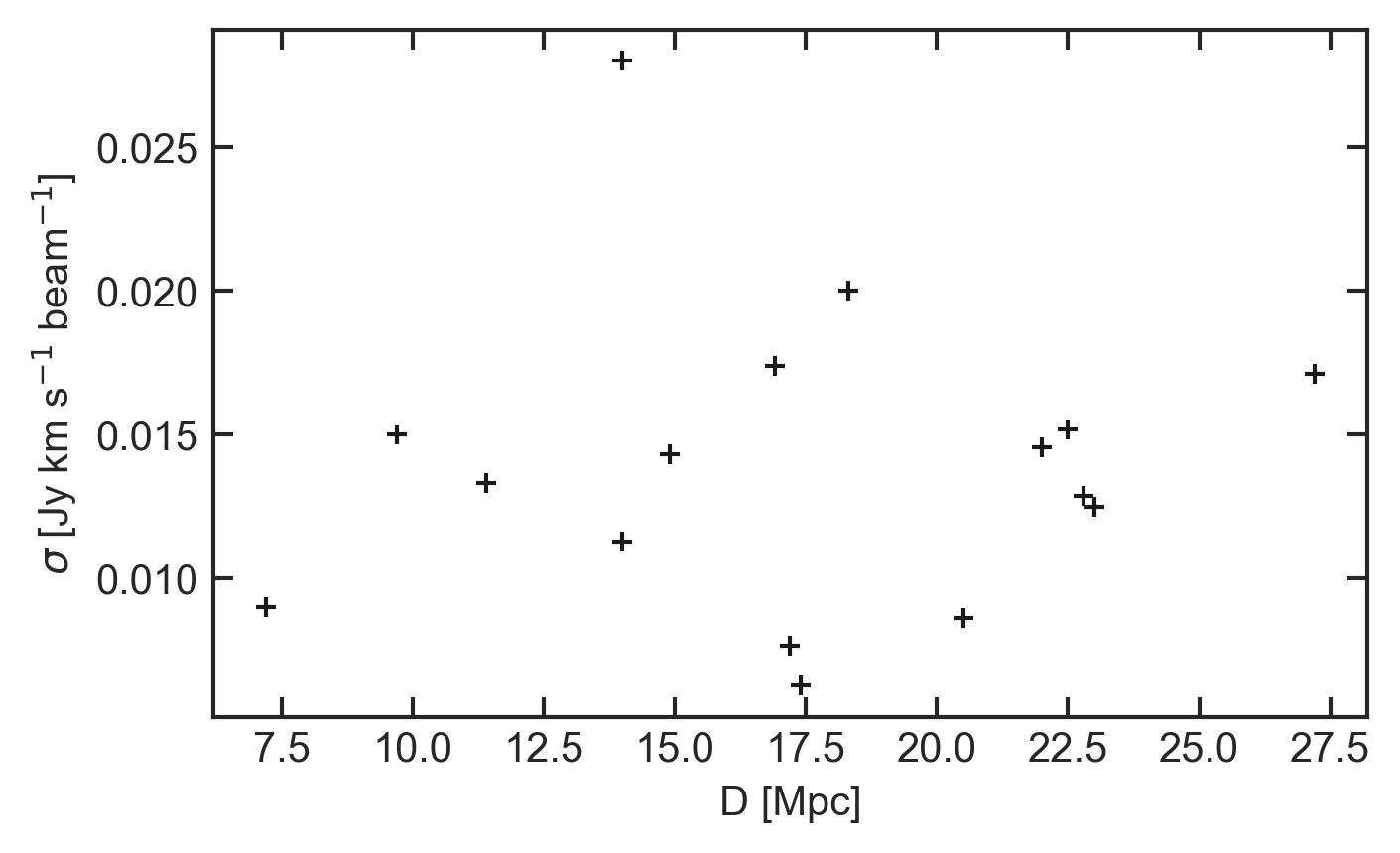}
\caption{
RMS sensitivities ($\sigma$) of the CO($3-2$) maps 
in units of Jy km s$^{-1}$ beam$^{-1}$ as a
function of luminosity distance (in Mpc). 
No correlation is found between sensitivity and distance.
}
\label{fig:noise_jy}
\end{figure}

\begin{figure}[b]
\centering
\includegraphics[width=0.49\textwidth]{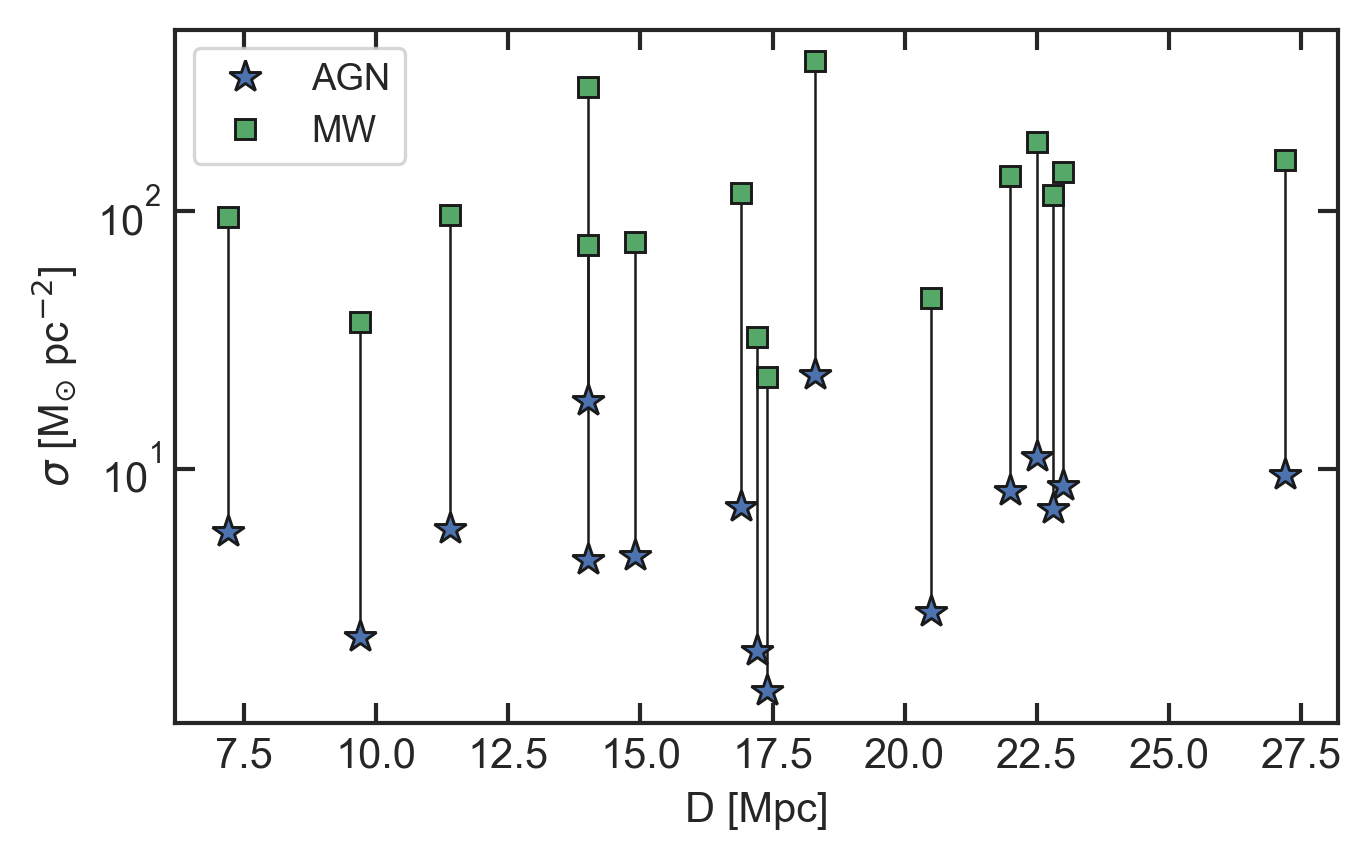}
\caption{
RMS sensitivities ($\sigma$) of the CO($3-2$) maps 
converted into molecular mass surface densities 
(in M$_{\odot}$ pc$^{-2}$) as a function of 
luminosity distance (in Mpc). 
For each galaxy we show two values, 
connected by a vertical bar: 
green squares correspond to the case $r_{31}=0.7$ 
\citep[average value in galaxy centers,][]{israel20} 
combined with the standard Milky Way disk 
conversion factor 
$X_{\text{CO}} = 2 \times 10^{20}$ 
mol cm$^{-2}$ (K km s$^{-1}$)$^{-1}$ \citep{bolatto13}; 
blue stars correspond to the case $r_{31}=2.9$ 
\citep[measured at the AGN position in NGC 1068,][]{viti14}
combined with the recommended conversion factor 
for the nuclear region of the Milky Way
$X_{\text{CO}} = 0.5 \times 10^{20}$ 
mol cm$^{-2}$ (K km s$^{-1}$)$^{-1}$ \citep{bolatto13}.
}
\label{fig:noise_msun}
\end{figure}

\begin{figure*}
\centering
\includegraphics[width=0.98\textwidth]{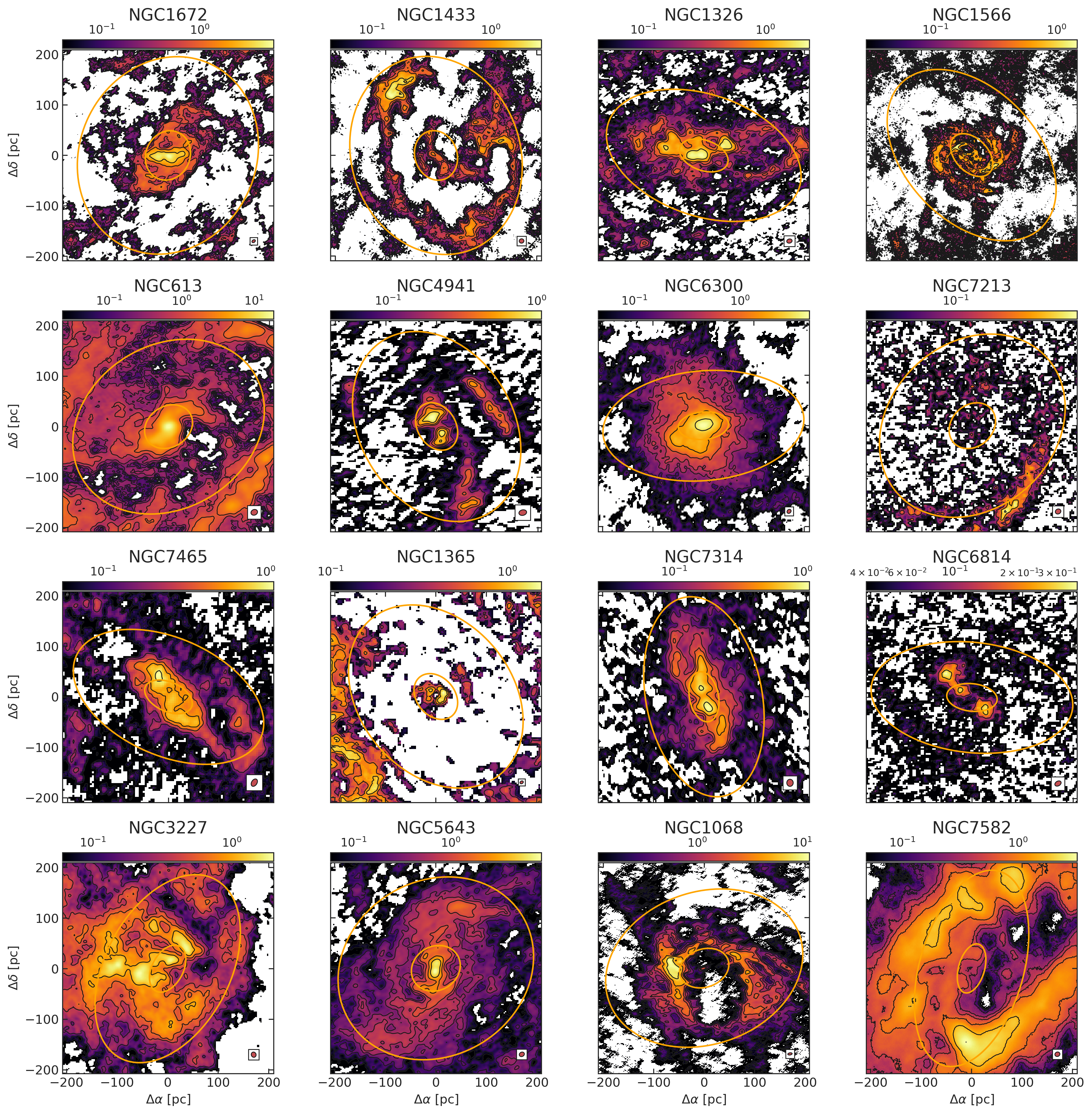}
\caption{
CO($3-2$) images of the sample galaxies.
Each image covers a region of 400 pc $\times$ 400 pc,
as in Figure~\ref{fig:example}.
The two ellipses in each panel correspond to circles 
with radii of 50 and 200 pc
projected onto the plane of the galaxy.
Galaxies are ordered by increasing $2-10$ keV 
luminosity, from top-left to bottom-right.
Contours correspond to 
$(3, 5, 10, 20, 30, 50, 100, 200) \times \sigma$.
The red ellipse inside the white square 
at the bottom-right corner of each panel
represents the ALMA beam.
}
\label{fig:sample}
\end{figure*}

Figure~\ref{fig:noise_jy} shows the RMS 
sensitivity (in Jy km s$^{-1}$ beam$^{-1}$) 
as a function of luminosity distance, 
demonstrating that there is no correlation 
between the two. 
Figure~\ref{fig:noise_msun} 
presents the same sensitivities 
converted into molecular mass surface densities 
(M$_\odot$ pc$^{-2}$), 
using the conversion factors discussed in 
Section~\ref{sec:sample}.
In addition, we
show the ALMA CO($3-2$) integrated 
intensity maps (moment 0) of the sample 
(Figure~\ref{fig:sample}).

\begin{figure*}
\centering
\includegraphics[width=0.98\textwidth]{
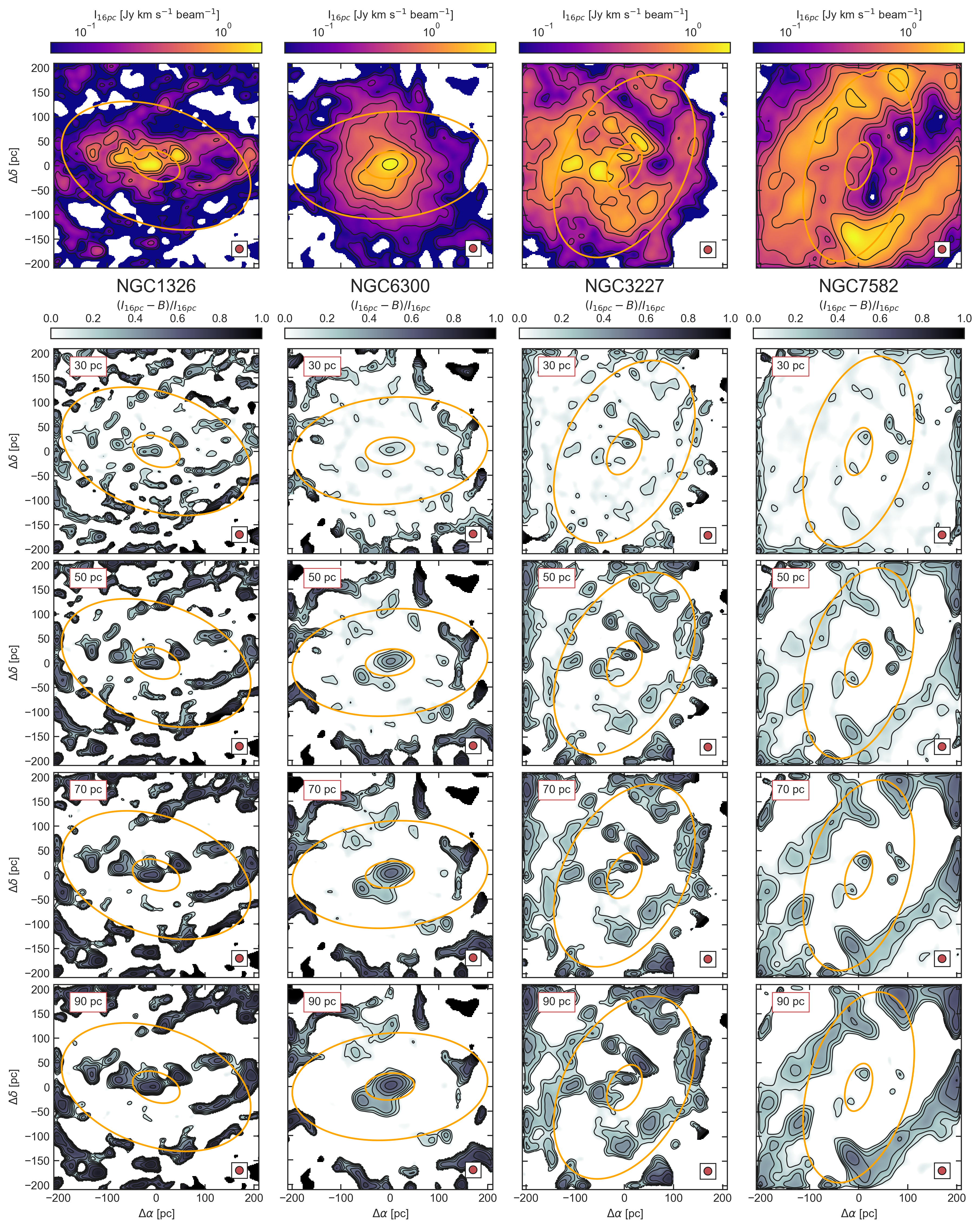}
\caption{
CO($3-2$) maps smoothed to a common physical
resolution of 16 pc for four representative
galaxies (\textit{top row}):
NGC 1326, NGC 6300, NGC 3227, and NGC 7582
(from left to right).
Contours are the same as in
Figure~\ref{fig:sample}.
The colourbars indicate flux in
Jy km s$^{-1}$ beam$^{-1}$.
From the second to the fifth row,
for each galaxy (each column),
we show the clumpiness map computed with
Method \#1, using Gaussian smoothing
kernels with FWHM of
30, 50, 70, and 90 pc
(from top to bottom).
Contours are plotted
every 0.1 between 0 and 1.
The two ellipses in each panel correspond to circles 
with radii of 50 and 200 pc
projected onto the plane of the galaxy.
The red ellipse inside the white square 
at the bottom-right corner of each panel
represents the ALMA beam.
}
\label{fig:mosaic4galaxies}
\end{figure*}

For four representative galaxies, 
we present the clumpiness maps computed at 
different smoothing scales 
(Figure~\ref{fig:mosaic4galaxies}).  
The galaxies are selected to cover
the $L_{\text{X}}$ range:
NGC 1326 ($L_{\text{X}} = 10^{39.9}$ erg s$^{-1}$),
NGC 6300 ($L_{\text{X}} = 10^{41.7}$ erg s$^{-1}$),
NGC 3227 ($L_{\text{X}} = 10^{42.4}$ erg s$^{-1}$), and
NGC 7582 ($L_{\text{X}} = 10^{43.5}$ erg s$^{-1}$).
These maps illustrate how the choice of 
smoothing kernel highlights structures 
of different sizes and contrast in the 
molecular gas distribution.

\begin{figure*}
\centering
\includegraphics[width=0.98\textwidth]{
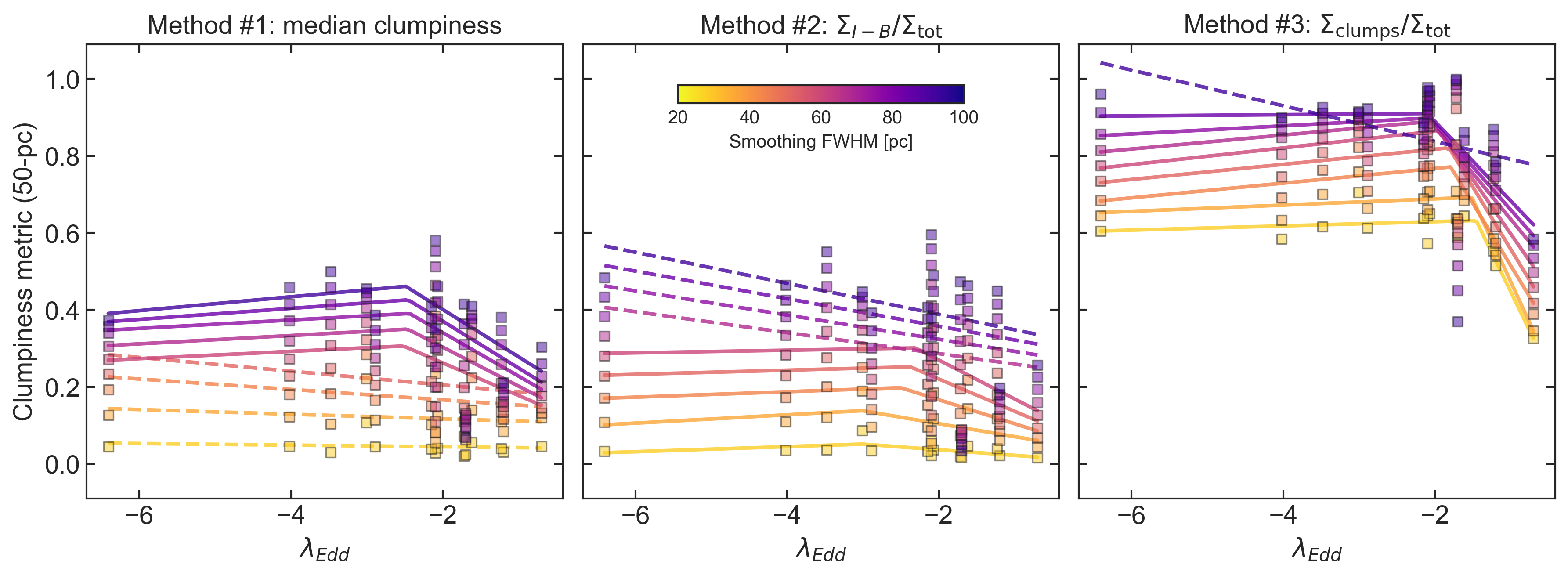}
\caption{
Clumpiness measured within the
50-pc-radius aperture, plotted
as a function of the AGN Eddington
ratio, $\lambda_{\text{Edd}}$.
Panels, lines, and markers are as in
Figure~\ref{fig:smooths_comparison}.
}
\label{fig:Edd_50pc}
\end{figure*}

\begin{figure*}
\centering
\includegraphics[width=0.98\textwidth]{
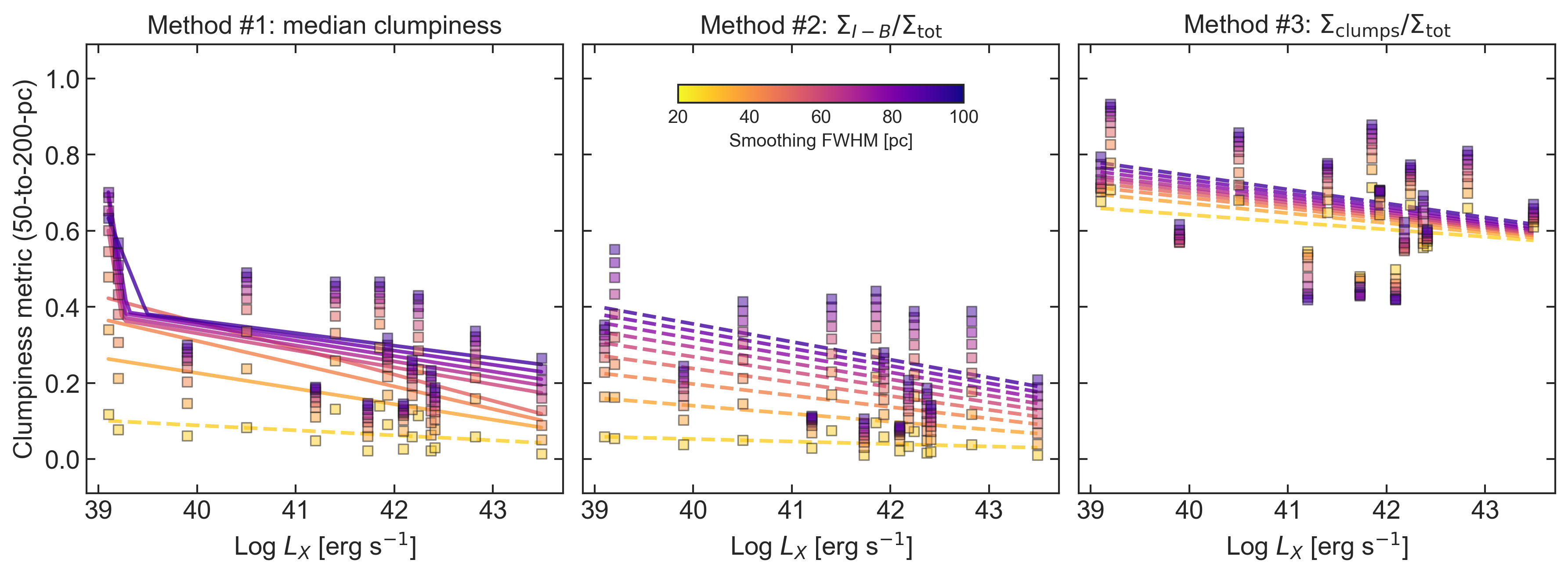}
\caption{
Clumpiness measured within the annular 
aperture between 50 and 200 pc, plotted
as a function of AGN X-ray luminosity.
Panels, lines, and markers are as in
Figure~\ref{fig:smooths_comparison}.
}
\label{fig:annular_smooths}
\end{figure*}

\begin{figure*}
\centering
\includegraphics[width=0.98\textwidth]{
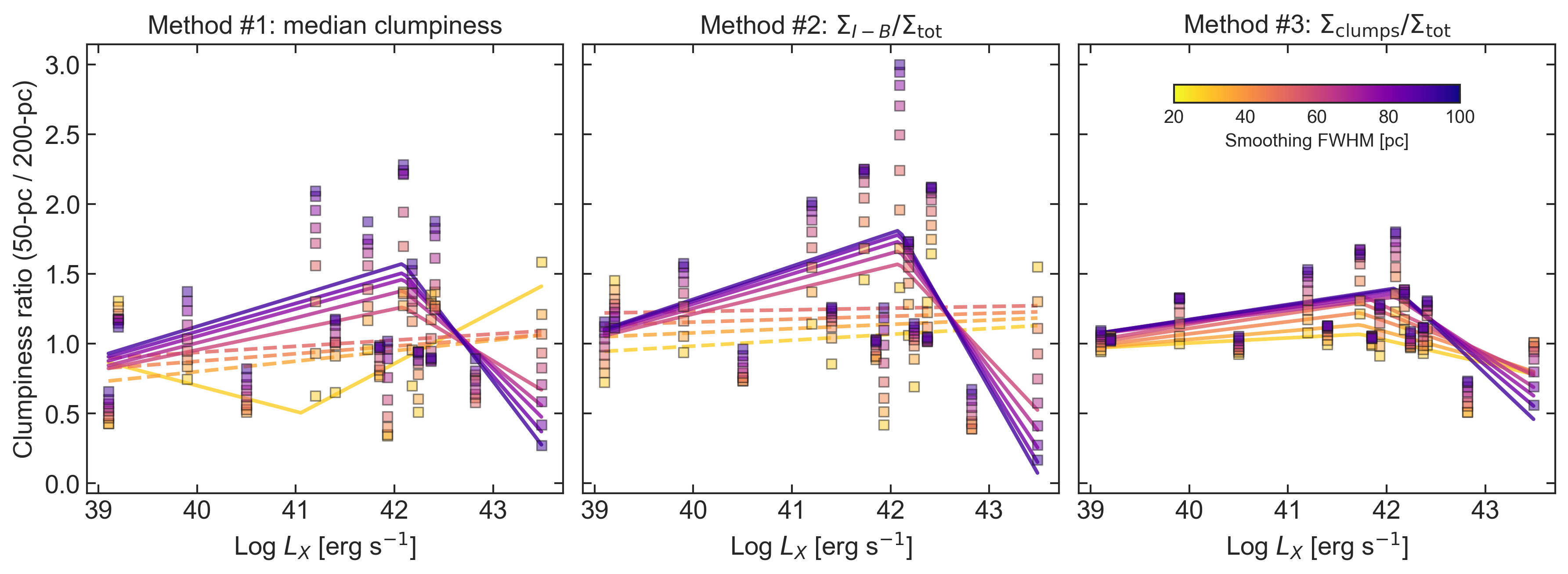}
\caption{
Ratio of clumpiness measured within the 
50-pc-radius aperture to that measured 
within the 200-pc-radius aperture, 
plotted as a function of AGN X-ray luminosity.
Panels, lines, and markers are as in
Figure~\ref{fig:smooths_comparison}.
}
\label{fig:RoR200pc_smooths}
\end{figure*}

\begin{figure*}
\centering
\includegraphics[width=0.98\textwidth]{
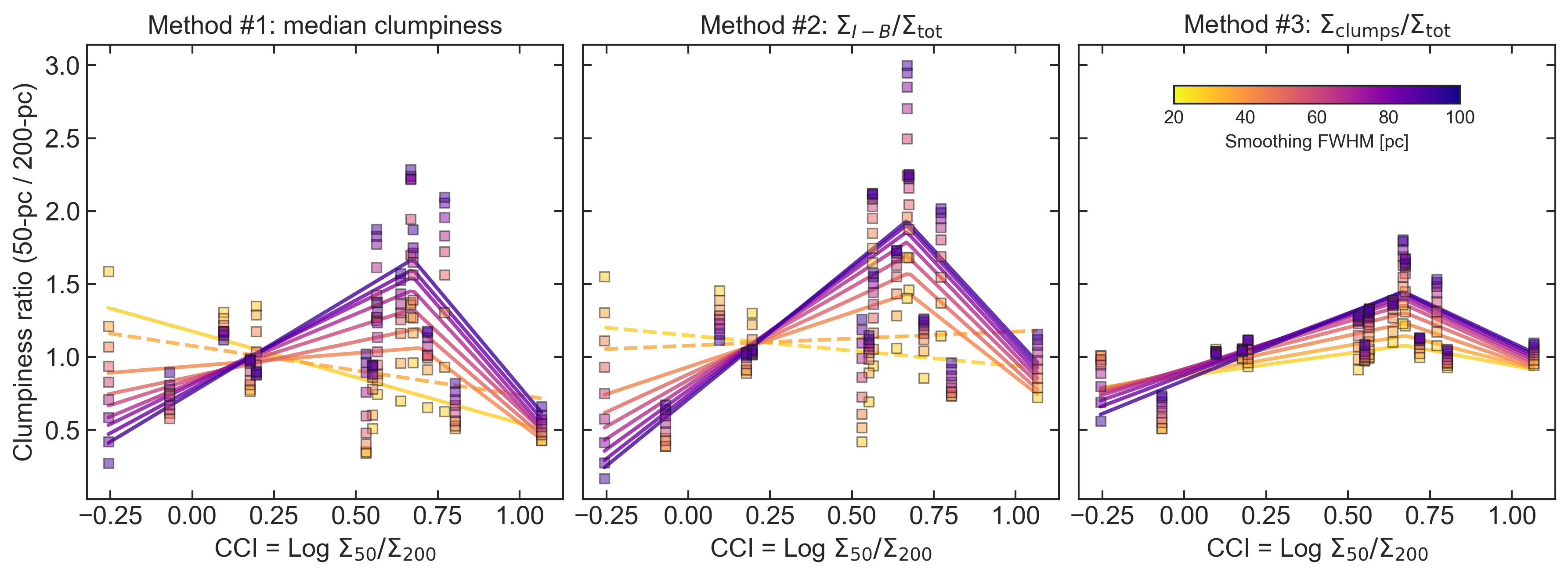}
\caption{
Ratio of clumpiness measured within the 
50-pc-radius aperture to that within the 
200-pc-radius aperture, plotted as a 
function of the cold molecular gas 
concentration index,
CCI $\equiv \log(\Sigma_{50} / \Sigma_{200})$.
Panels, lines, and markers are as in
Figure~\ref{fig:smooths_comparison}.
}
\label{fig:RoR200pc_CCI}
\end{figure*}

We also present additional tests 
and correlations to complement the main
results described in 
Section~\ref{sec:results}.
Specifically, we explore how clumpiness 
varies with the Eddington ratio 
(Figure~\ref{fig:Edd_50pc}),
when measured in an annular 
region excluding the nucleus 
(Figure~\ref{fig:annular_smooths}),
and how the ratio of nuclear to 
circumnuclear clumpiness depends on AGN 
luminosity and molecular gas concentration 
(Figures~\ref{fig:RoR200pc_smooths} 
and \ref{fig:RoR200pc_CCI}).

\begin{table}[b]
\centering
\caption{Properties of the sample.}
\label{tab:sample}
\begin{threeparttable}
\begin{tabular}{lccccc}
\toprule
Name & $D_{\text{L}}$ 
& $\log L_{\rm X}^{2-10 \text{ keV}}$ & $\log \lambda_{\rm Edd}$ 
& $i$ & PA \\
\midrule
{} & Mpc & erg s$^{-1}$ & {} & $^{\circ}$ & $^{\circ}$ \\
\midrule
NGC 613  & 17.2 & 41.20 & -3.48 & 36 & 122 \\
NGC 1068 & 14.0 & 42.82 & -0.70 & 41 & 289 \\
NGC 1326 & 14.9 & 39.90 & -4.02 & 53 &  71 \\
NGC 1365 & 18.3 & 42.09 & -2.15 & 41 &  40 \\
NGC 1433 &  9.7 & 39.20 &    -- & 33 & 199 \\
NGC 1566 &  7.2 & 40.50 & -2.89 & 49 &  44 \\
NGC 1672 & 11.4 & 39.10 & -6.41 & 29 & 155 \\
NGC 3227 & 23.0 & 42.37 & -1.20 & 52 & 152 \\
NGC 4941 & 20.5 & 41.40 & -2.10 & 41 & 212 \\
NGC 5643 & 16.9 & 42.41 & -1.23 & 30 & 301 \\
NGC 6300 & 14.0 & 41.73 & -1.72 & 57 &  95 \\
NGC 6814 & 22.8 & 42.24 & -1.62 & 57 &  84 \\
NGC 7213 & 22.0 & 41.85 & -3.01 & 35 & 133 \\
NGC 7314 & 17.4 & 42.18 & -2.07 & 55 & 191 \\
NGC 7465 & 27.2 & 41.93 & -2.10 & 54 &  66 \\
NGC 7582 & 22.5 & 43.49 & -1.70 & 59 & 344 \\
\bottomrule
\end{tabular}
\begin{tablenotes}
\item \textbf{Notes.}
Distances ($D_{\text{L}}$) are median values
of redshift-independent distances taken 
from the Nasa Extragalactic Database (NED).
X-ray luminosities ($L_{\rm X}^{2-10 \text{ keV}}$)
are taken from \cite{ricci17}; \cite{combes19}; 
\cite{garciaburillo24}.
Eddington ratios ($\lambda_{\rm Edd}$)
are taken from \cite{koss17};
\cite{garciaburillo24}.
Inclinations ($i$) and position angles (PA)
were derived with the
\texttt{kinemetry} software \citep{krajnovic06}
as presented in \cite{garciaburillo21}.
\end{tablenotes}
\end{threeparttable}
\end{table}

\end{appendix}

\end{document}